\documentclass[fleqn,usenatbib]{mnras}

\usepackage{newtxtext,newtxmath}

\usepackage[T1]{fontenc}

\DeclareRobustCommand{\VAN}[3]{#2}
\let\VANthebibliography\thebibliography
\def\thebibliography{\DeclareRobustCommand{\VAN}[3]{##3}\VANthebibliography}

\usepackage[dvipsnames]{xcolor}

\usepackage{graphicx}	
\usepackage{amsmath}	

\newcommand*{\possis}{\textsc{possis}}

\newcommand{\revised}[1]{#1}

\title[POSSIS 2.0]{The critical role of nuclear heating rates, thermalization efficiencies and opacities for kilonova modelling and parameter inference}

\author[M. Bulla]{
Mattia Bulla$^{1,2,3}$\thanks{E-mail: mattia.bulla@unife.it}
\\
$^{1}$Department of Physics and Earth Science, University of Ferrara, via Saragat 1, I-44122 Ferrara, Italy\\
$^{2}$INFN, Sezione di Ferrara, via Saragat 1, I-44122 Ferrara, Italy\\
$^{3}$INAF, Osservatorio Astronomico d’Abruzzo, via Mentore Maggini snc, 64100 Teramo, Italy\\
}

\date{Accepted 2023 January 18. Received 2023 January 9; in original form 2022 November 30}

\pubyear{2015}

\begin{document}
\label{firstpage}
\pagerange{\pageref{firstpage}--\pageref{lastpage}}
\maketitle

\begin{abstract}
We present an improved version of the 3D Monte Carlo radiative transfer code \possis~to model kilonovae from neutron star mergers, wherein nuclear heating rates, thermalization efficiencies and wavelength-dependent opacities depend on local properties of the ejecta and time. Using an axially-symmetric two-component ejecta model, we explore how simplistic assumptions on heating rates, thermalization efficiencies and opacities often found in the literature affect kilonova spectra and light curves. Specifically, we compute five models: one (\texttt{FIDUCIAL}) with an appropriate treatment of these three quantities, one (\texttt{SIMPLE-HEAT}) with uniform heating rates throughout the ejecta, one (\texttt{SIMPLE-THERM}) with a constant and uniform thermalization efficiency, one (\texttt{SIMPLE-OPAC}) with grey opacities and one (\texttt{SIMPLE-ALL}) with all these three simplistic assumptions combined. We find that deviations from the \texttt{FIDUCIAL} model are of several ($\sim1-10$) magnitudes and are generally larger for the \texttt{SIMPLE-OPAC} and \texttt{SIMPLE-ALL} compared to the \texttt{SIMPLE-THERM} and \texttt{SIMPLE-HEAT} models. The discrepancies generally increase from a face-on to an edge-on view of the system, from early to late epochs and from infrared to ultraviolet/optical wavelengths. Our work indicates that kilonova studies using either of these simplistic assumptions ought to be treated with caution and that appropriate systematic uncertainties ought to be added to kilonova light curves when performing inference on ejecta parameters.
\end{abstract}




\section{Introduction}

The detection of an electromagnetic  counterpart to the gravitational-wave (GW) event GW170817 \citep{Abbott2017a} marked year zero of the multi-messenger GW era. This event was generated by the merger of two neutron stars and gave rise to multiple transients detected across the whole electromagnetic spectrum by ground-based and spaced-borne instruments \citep{Abbott2017c}: a short gamma-ray burst (GRB) from a relativistic jet \citep[GRB170817A;][]{Goldstein2017,Savchenko17,Abbott2017d}, the associated afterglow from the interaction of the jet with the circumburst medium \citep[e.g.][]{Alexander2017,Haggard2017,Hallinan2017,Margutti2017,Troja2017,DAvanzo2018}, a ``kilonova'' powered by the radioactive decay of $r-$process nuclei forged during the merger \citep[AT\,2017gfo; e.g.][]{Andreoni2017,Arcavi2017, Cowperthwaite2017,Drout2017,Evans2017,Kasliwal2017,Mccully2017, Pian2017,Smartt2017,SoaresSantos2017,Utsumi2017} and possibly a ``kilonova afterglow'' from the late interaction of merger ejecta with the circumburst medium \citep[][but see \citealt{Balasubramanian2022} and \citealt{Troja2022}]{Hajela2022}.

Despite the reasonable good agreement between AT\,2017gfo and existing kilonova models, inferred properties of the ejecta like masses and velocities \citep[e.g.][]{Chornock2017,Cowperthwaite2017,Kasen2017,Kasliwal2017,Smartt2017,Tanaka2017,Villar2017,Coughlin2018,Breschi2021,Ristic2022} were found to be in tension with those predicted by numerical relativity simulations (see fig. 1 in \citealt{Siegel2019} or fig. 6 in \citealt{Nedora2021} for a detailed comparison with models). However, the ejecta parameters were inferred using either semi-analytical, one-dimensional or combination of one-dimensional models, therefore overlooking important aspects like the observer viewing angle, the geometry of the ejecta and the reprocessing of radiation between different ejecta components. These aspects are naturally incorporated in studies performing multi-dimensional radiative transfer simulations for kilonovae \citep[e.g.][]{Wollaeger2018,Bulla2019b,Darbha2020,Kawaguchi2018,Korobkin2021,Collins2022}. Indeed, some of these works suggest that the claimed tension is alleviated when multi-dimensional aspects and radiation transport are properly taken into account \citep{Kawaguchi2018,Kawaguchi2020,Bulla2019b,Dietrich2020,Kedia2022}.

The viewing angle, the ejecta geometry and the reprocessing of radiation are not the only aspects that are known to affect kilonova spectra and light curves. In fact, key ingredients for kilonova modelling include nuclear heating rates \citep{Roberts2011,Lippuner2015,Wu2019,Rosswog2022}, thermalization efficiencies \citep{Metzger2010,Barnes2016,Hotokezaka2016} and opacities \citep{Barnes2013,Kasen2013,Tanaka2013,Metzger2014}. These three quantities determine, respectively, the amount of energy coming from radioactive decay of $r-$process nuclei, what fraction of this energy thermalizes within the ejecta and leads to kilonova radiation, and finally how kilonova radiation interacts with matter while propagating through the expanding material. In this work, we study the critical role of these three quantities in shaping the kilonova spectra and light curves. To this aim, we use a new version of the 3D Monte Carlo radiative transfer code \possis~in which heating rates, thermalization efficiencies and opacities depend on local properties of the ejecta and investigate how simplistic assumptions on either (or all) of these quantities affect kilonova observables and bias parameter inference. The purpose of this study is not to explore how kilonova observables are affected by uncertainties on these quantities as done elsewhere \citep[see e.g.][]{Barnes2021,Zhu2021,Pognan2022}, but rather how they are biased by simplistic assumptions often found in the literature, e.g. uniform heating rates, constant and uniform thermalization efficiencies and grey opacities. 

The paper is organized as follows. In Section~\ref{sec:possis}, we outline the upgrades to \possis~compared to the version in \citet{Bulla2019b}. In Section~\ref{sec:models}, we then discuss the ejecta model assumed in this work and the different simulations performed. In Section~\ref{sec:results}, we present results of our simulations, before summarizing and concluding in Section~\ref{sec:conclusions}.

\section{POSSIS 2.0}
\label{sec:possis}

\possis~is a 3D Monte Carlo radiative transfer code that models flux and polarization spectral time series for explosive transients \citep{Bulla2019b}. The code has been used mainly to predict kilonova observables \citep[e.g.][]{Bulla2019a,Dietrich2020,Anand2021,Bulla2021} but also to study polarization signatures of supernovae \citep{Inserra2016} and tidal disruption events \citep[TDEs;][Charalampopoulos et al., submitted]{Leloudas2022}. The code simulates the propagation of Monte Carlo photon packets in an homologously expanding medium and can model the interaction of packets with matter via electron scattering, bound-bound, bound-free and free-free transitions, where opacities for each of these processes are taken as an input. Properties of the photon packets (direction of propagation, energy, frequency and Stokes parameters) are updated after every interaction and the final synthetic observables (flux and polarization spectra) are computed using the ``virtual'' packet EBT technique described in \citet{Bulla2015} rather than using the standard angular binning of escaping packets that is subject to a higher numerical noise. We refer the reader to \citet{Bulla2015} and \citet{Bulla2019b} for more details about the code. The code has developed significantly from its first release and now features a better treatment of the ejecta temperature, nuclear heating rates, thermalization efficiencies and opacities, which are detailed in the next sections. 

\subsection{Temperature}
\label{sec:temp}

In \citet{Bulla2019b}, the ejecta temperature was assumed to be uniform throughout the ejecta and its time dependence parameterized by a simple power law $T\propto t^{-\alpha}$, with the index $\alpha=0.4$ found to give reasonable fits to AT\,2017gfo. In the new version of the code, instead, the ejecta temperature is computed assuming perfect coupling between matter and radiation and computing the radiation temperature $T_{\rm R}$ as detailed below and implemented in other supernova and kilonova radiative transfer codes \citep[e.g.][]{Lucy2005,Kromer2009,Tanaka2013,Magee2018}. This assumption is likely to break down at late times, at which point the ejecta temperature are likely to be underestimated. We note that the latest kilonova grids performed with \possis~for both binary neutron star \cite[e.g.][]{Dietrich2020,Perez2022} and neutron-star black-hole \citep{Anand2021} mergers adopt this new temperature treatment.

At the start of the simulation, $t_i$, the temperature is initialized in each cell using the local heating from the radioactive decay of $r-$process nuclei. Specifically, the radiation energy density $U_{\rm R}$ is calculated as 
\begin{equation}
   U_{\rm R} = \int^{t_i}_0 \rho(t)\,\dot{\epsilon}(t)\,\epsilon_{\rm th}\,\bigg(\frac{t}{t_i}\bigg)\,dt~~~~,
\end{equation}
where $\rho$, $\dot{\epsilon}$ and $\epsilon_{\rm th}$ are the mass density, heating rates (Section~\ref{sec:heat}) and thermalization efficiencies (Section~\ref{sec:therm}) in the given cell, respectively, \revised{the term $t/t_i$ accounts for adiabatic energy losses} and the integral is performed from the time of merger $t=0$ to $t=t_i$. The initial temperature $T_{\rm R}$ is then computed assuming that the mean intensity of the radiation field
\begin{equation}
    <J>\,= \frac{U_{\rm R} c}{4\pi} 
\end{equation}
follows the Stefan-Boltzmann law, i.e.
\begin{equation}
\label{eq:Tr}
    T_{\rm R} = \bigg(\frac{\pi \,<J>}{\sigma}\bigg)^{1/4} = \bigg(\frac{U_{\rm R} c}{4\sigma}\bigg)^{1/4}
\end{equation}
where $c$ is the speed of light and $\sigma$ is the Stephan-Boltzmann constant. During the simulation, the temperature is then updated at the end of each time-step using Equation~(\ref{eq:Tr}) and Monte Carlo estimators \citep{Mazzali1993,Lucy2003} for the mean intensity,
\begin{equation}
    <J>\,= \frac{1}{4\pi\Delta t\, V} \sum{e_{\rm cmf}\,\Delta l}~~~,
\end{equation}
where $\Delta t$ is the time-step duration, $V$ is the cell volume and the sum is performed over all the photon packets with comoving-frame energy $e_{\rm cmf}$ travelling a path $\Delta l$ through the given cell and in the given time-step.

\subsection{Nuclear heating rates}
\label{sec:heat}

For nuclear heating rates, \citet{Bulla2019b} adopted the following analytic formula from \citet{Korobkin2012}
\begin{equation}
    \label{eq:heat}
    \dot{\epsilon}(t)=\epsilon_0\,\bigg(\frac{1}{2}-\frac{1}{\pi}\,{\rm arctan}\,\frac{t-t_0}{\sigma}\bigg)^\alpha
\end{equation}
where $\epsilon_0=2\times10^{18}$ erg s$^{-1}$ g$^{-1}$, $t_0=1.3$\,s, $\sigma=0.11$\,s and $\alpha=1.3$. This formula captures the plateau phase found in the first $\sim1$s after the merger and the subsequent power-law decay with an index of $\alpha=1.3$. The heating rates were assumed by \citet{Bulla2019b} to be uniform throughout the ejecta although a dependence on local properties like velocity and electron fraction is expected and the formula in Eq.~\ref{eq:heat} was constructed from numerical simulations of dynamical ejecta with lanthanide-rich compositions \citep{Korobkin2012}. In contrast, the new version of \possis~implements heating rate libraries from \citet{Rosswog2022} and specifically an improved fitting formula
\begin{equation}
    \begin{aligned}
    \dot{\epsilon}(t)=\epsilon_0\,\bigg(\frac{1}{2}-\frac{1}{\pi}\,{\rm arctan}\,\frac{t-t_0}{\sigma}\bigg)^\alpha\,\bigg(\frac{1}{2}+\frac{1}{\pi}\,{\rm arctan}\,\frac{t-t_1}{\sigma_1}\bigg)^{\alpha_1}+\\
    +\,C_1\,e^{-t/\tau_1}+\,C_2\,e^{-t/\tau_2}+\,C_3\,e^{-t/\tau_3}~~~~,
    \end{aligned}
\end{equation}
where the different coefficients are no longer uniform as in \citet{Korobkin2012} but rather depend on local values of velocities $\varv$ and electron fraction $Y_e$ within the ejecta (see their eq. 2 and table 1). The heating rate library this formula was fitted to was computed using the Winnet network \citep{Winteler2012} as described in details by \citet{Rosswog2022}. 

\subsection{Thermalization efficiencies}
\label{sec:therm}

\begin{table*}
    \centering
    \begin{tabular}{l|c|c|c}
          Model & Heating rates & Thermalization efficiencies & Opacities \\
          & [erg s$^{-1}$ g$^{-1}$] &  & [cm$^2$ g$^{-1}$] \\
         \hline
         \texttt{FIDUCIAL} & \textcolor{OliveGreen}{$\dot{\epsilon}(t,\varv,Y_{\rm e})$} & \textcolor{OliveGreen}{$\epsilon_{\rm th}(t,\rho)$} & \textcolor{OliveGreen}{$\kappa_{\rm sc}(t,\rho,T,Y_{\rm e})$ + $\kappa_{\rm bb}(\lambda,t,\rho,T,Y_{\rm e})$} \\ 
         \texttt{SIMPLE-HEAT} & \textcolor{Red}{$\dot{\epsilon}(t)$\,=\,uniform} & \textcolor{OliveGreen}{$\epsilon_{\rm th}(t,\rho)$} & \textcolor{OliveGreen}{$\kappa_{\rm sc}(t,\rho,T,Y_{\rm e})$ + $\kappa_{\rm bb}(\lambda,t,\rho,T,Y_{\rm e})$} \\ 
         \texttt{SIMPLE-THERM} & \textcolor{OliveGreen}{$\dot{\epsilon}(t,\varv,Y_{\rm e})$} & \textcolor{Red}{$\epsilon_{\rm th}=0.5$} & \textcolor{OliveGreen}{$\kappa_{\rm sc}(t,\rho,T,Y_{\rm e})$ + $\kappa_{\rm bb}(\lambda,t,\rho,T,Y_{\rm e})$} \\ 
         \texttt{SIMPLE-OPAC} & \textcolor{OliveGreen}{$\dot{\epsilon}(t,\varv,Y_{\rm e})$} & \textcolor{OliveGreen}{$\epsilon_{\rm th}(t,\rho)$} & \textcolor{Red}{$\kappa_{\rm dyn,lf}=0.5$\,|\,$\kappa_{\rm dyn,lr}=10$\,|\,$\kappa_{\rm wind}=3$} \\ 
         \texttt{SIMPLE-ALL} & \textcolor{Red}{$\dot{\epsilon}(t)$\,=\,uniform} & \textcolor{Red}{$\epsilon_{\rm th}=0.5$} & \textcolor{Red}{$\kappa_{\rm dyn,lf}=0.5$\,|\,$\kappa_{\rm dyn,lr}=10$\,|\,$\kappa_{\rm wind}=3$} \\ 
         \hline
    \end{tabular}
    \caption{Summary of the assumed nuclear heating rates, thermalization efficiencies and opacities for the five models simulated. Assumptions that depend on local properties of the ejecta are highlighted in green, while those that do not are shown in red. The heating rates $\dot{\epsilon}(t,\varv,Y_{\rm e})$ are from \citet{Rosswog2022}, the thermalization efficiencies $\epsilon_{\rm th}(t,\rho)$ are computed following \citet{Barnes2016} and \citet{Wollaeger2018} and the electron-scattering $\kappa_{\rm sc}(t,\rho,T,Y_{\rm e})$ and bound-bound $\kappa_{\rm bb}(\lambda,t,\rho,T,Y_{\rm e})$ opacities are from \citet{Tanaka2020}. See Section~\ref{sec:possis} and Section~\ref{sec:sim} for more details. }
    \label{tab:models}
\end{table*}

\citet{Bulla2019b} adopted a uniform and constant value of $\epsilon_{\rm therm}=0.5$, i.e. half of the energy from radioactive decay was assumed to thermalize within the ejecta and lead to kilonova emission. In contrast, the new version of the code presented here computes density-dependent thermalization efficiencies from \citet{Wollaeger2018} that follow the methodology from \citet{Barnes2016}. First, we compute what fraction $f_j$ of the heating rates discussed in Section~\ref{sec:heat} are carried away by each species $j$ ($\alpha-$ and $\beta-$particles, fission fragments, gamma rays and neutrinos). Then, we compute the thermalization efficiencies for each species. For $\alpha-$ and $\beta-$particles, the local thermalization efficiencies at time $t$ and location ${\bf r}$ are calculated as
\begin{equation}
    \epsilon_{\rm therm}^{\,j}(t,{\bf r}) = \frac{\ln\Big(1+\frac{2\,A_j}{t\,\rho(t,{\rm r})} \Big)}{1+\frac{2\,A_j}{t\,\rho(t,{\rm r})}}~~~~,
\end{equation}
where $[A_\alpha,A_\beta,A_{ff}]=[1.2,1.3,0.2]\times10^{-11}$ g cm$^{-3}$ s. For $\gamma-$rays, instead, the thermalization efficiciency is set to 
\begin{equation}
    \epsilon_{\rm therm}^{\gamma}(t,{\bf r}) = 1-e^{-\tau_\gamma}~~~~,
\end{equation}
where $\tau_\gamma(t,{\bf r})=\int^\infty_{\bf r}\kappa_\gamma\rho(t,{\bf r})\,d{\bf r}$ is the optical depth to the boundary computed within the code assuming an averaged gamma-ray opacity of $\kappa_\gamma=0.1$\,cm$^2$\,g$^{-1}$. As a result, the total thermalization efficiency used to multiply the heating rates from Section~\ref{sec:heat} is given by
\begin{equation}
    \epsilon_{\rm therm}(t,{\bf r})=\sum{f_j\times \epsilon_{\rm therm}^{\,j}}
\end{equation}
for $j$ in [$\alpha,\beta$, fission fragments,$\gamma$]. 

For the calculations performed in this work, we assume $f_\alpha=0.05$, $f_\beta=0.2$, $f_{\rm ff}=0$, $f_\gamma=0.4$ and $f_\nu=0.35$ (i.e. 35\% of the energy is lost through neutrinos) based on fig. 4 of \citet{Wollaeger2018}.

\subsection{Opacities}
\label{sec:opac}

Electron (Thomson) scattering and bound-bound transitions are the main source of opacities at the times and wavelengths relevant for kilonova emission \citep{Tanaka2020,Fontes2020,Fontes2023}. In \citet{Bulla2019b}, simple analytic functions were used to go beyond the grey-opacity assumption and model the time and wavelength dependence of electron scattering opacities. Simple power-law functions that mimic the detailed time- and wavelength-dependence from \citet{Tanaka2018} were adopted (see fig. 2 in \citealt{Bulla2019b}). Moreover, opacities were assumed to be uniform within a given ejecta region, with different values adopted in so-called ``lanthanide-poor'' and ``lanthanide-rich'' ejecta \citep[see][for a three-component model with a ``lanthanide-intermediate'' ejecta]{Dietrich2020}. In contrast, the new version of \possis~presented here implements time- and wavelength-dependent opacities from \citet{Tanaka2020} that depend on local properties of the ejecta like the density $\rho$, the temperature $T=T_{\rm R}$ (see Section~\ref{sec:temp}) and the electron fraction $Y_{\rm e}$. These opacities were calculated assuming Local Thermodynamic Equilibrium (LTE) and using the so-called expansion opacities formalism \citep{Karp1977,Eastman1993}. 

Opacities from \citet{Tanaka2020} are tabulated at 
times $t=[0.5,1,2,3,..,19,20]$\,d after the merger, wavelengths $(\lambda_1,\lambda_2,\Delta\lambda)=(534.50,35\,000,34.5)$\,\AA, densities $(\log\rho_1,\log\rho_2,\Delta\log\rho)=(-19.5,-5.0,0.5)$, temperature $(T_1,T_2,\Delta T)=(1000,50\,500,500)$ and electron fraction $(Y_{\rm e,1},Y_{\rm e,2},\Delta Y_{\rm e})=(0.1,0.4,0.05)$. This corresponds to $4.4\times10^5$ values for the electron scattering opacities $\kappa_{\rm sc}(t,\rho,T,Y_{\rm e})$ and $4.4\times10^8$ values for bound-bound opacities $\kappa_{\rm bb}(\lambda,t,\rho,T,Y_{\rm e})$. To reduce the memory footprint of the simulations, opacities are rebinned of a factor of 8 in wavelength ($\Delta\lambda_{\rm new}=276$\,\AA). Although a rebinning washes out information on opacity below $\sim50$\,\AA, we note that atomic calculations from \citet{Tanaka2020} are not calibrated with experiments and therefore have only a $\sim$\,20\% accuracy on the transition wavelengths (Tanaka, private communication). During a simulation with the new version of \possis, ejecta properties ($\rho$, $T$ and $Y_{\rm e}$), time and comoving frequency/wavelength of the photon packet determine the opacities needed to model the radiative transfer. We note that the closest opacities values in the pre-tabulated grid from \citet{Tanaka2020} are currently selected, although we plan to adopt an interpolation scheme in the future. 

The opacities from \citet{Tanaka2020} are computed for neutral and singly to triply ionized stages (I$-$IV). As a consequence, bound-bound opacities are likely underestimated at times earlier than $\sim0.5-1$\,days from the merger, when the ejecta are hot ($T\gtrsim 20\,000$ K) and more highly ionized (see \citealt{Banerjee2020} and \citealt{Banerjee2022} for updated opacities grids up to ionization stage XI). Therefore, we neglect epochs at $t\lesssim0.5$\,d in the rest of the paper.


\section{Simulations}
\label{sec:models}

\begin{figure}
\centering
\includegraphics[width=1\columnwidth]{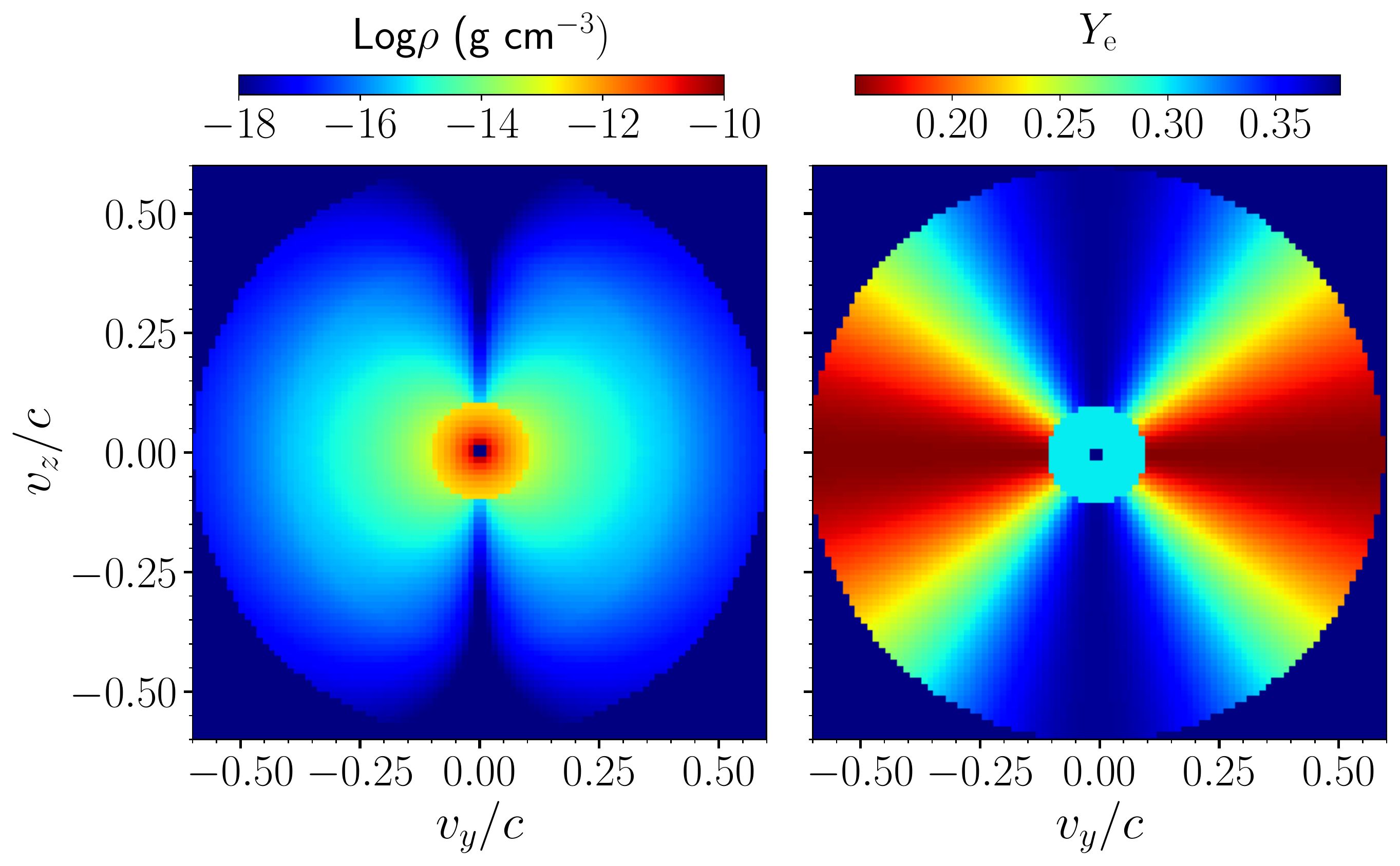}
\caption{Density (left) and $Y_{\rm e}$ (right) maps in the $\varv_y-\varv_z$ velocity plane for the ejecta model used in this work. A spherical disk-wind ejecta component extends from 0.02 to 0.1c, while a dynamical-ejecta component extends from 0.1 to 0.6c. The model is axially symmetric about the $\varv_z$ axis and the merger plane is defined by the $\varv_x-\varv_y$ plane. Densities are relative to 1~day after the merger. }
\label{fig:ejecta}
\end{figure}

In the following, we discuss details of the ejecta model assumed (Section~\ref{sec:ejecta}) and of the radiative transfer simulations performed (Section~\ref{sec:sim}).

\subsection{Input ejecta model}
\label{sec:ejecta}

We predict kilonova observables for an axially-symmetric two-component model with ejecta properties that are consistent with those found in numerical-relativity simulations of binary-neutron star mergers \cite[e.g.][]{Radice2018,Nedora2021}. A first component represents post-merger disk-wind ejecta, which are assumed to be spherical and to extend from a minimum velocity of 0.02c to a maximum velocity of 0.1c (mass-weigthed averaged velocity of $\bar{\varv}_{\rm wind}=0.05$c). The mass of the disk-wind component is set to $m_{\rm wind}=0.05\,M_\odot$ and the composition assumed to be relatively lanthanide-poor ($Y_{\rm e,wind}=0.3$). A second component represents the dynamical ejecta, which extend from a minimum velocity of 0.1c to a maximum velocity of 0.6c (mass-weigthed averaged velocity of $\bar{\varv}_{\rm dyn}=0.2$c). The total mass of the dynamical ejecta is set to $m_{\rm dyn}=0.005\,M_\odot$ (10 times smaller than the disk-wind ejecta mass) and the composition assumed to vary from lanthanide-rich close to the merger plane to lanthanide-poor at polar angles. Specifically, we follow \citet{Setzer2022} and adopt the following prescription for the angular dependence of the electron fraction that captures reasonably well the one found in numerical-relativity simulations \citep{Radice2018}:
\begin{equation}
    \label{eq:ye}
    Y_{\rm e, dyn}(\theta) = a\,\cos^2(\theta)\,+\,b~~~, 
\end{equation}
where $\theta$ is the polar angle and $a$ and $b$ are coefficients set to give a desired average $\bar{Y}_{\rm e, dyn}$. For this study, we set $a=0.2195$ and $b=0.1561$ which corresponds to $\bar{Y}_{\rm e, dyn}=0.2$ and an electron fraction varying from a minimum of $Y_{\rm e, dyn-min}=b\sim0.16$ ($\theta=90^\circ$ in Equation~\ref{eq:ye}) in the merger plane to a maximum of $Y_{\rm e, dyn-max}=(a+b)\sim0.38$ at the pole ($\theta=0$ in Equation~\ref{eq:ye}). This angular dependence leads to lanthanide-rich compositions ($Y_{\rm e}<0.25$) for ejecta within an half-opening angle $\phi=40^\circ$ from the merger plane. For the density profiles, we adopt the following analytic functions
\begin{equation}
    \rho(r,t) \propto \begin{cases} \,r^{-3}\,t^{-3} \,\,,\,\,\,\,0.02c\leq r/t \leq 0.1c\,\,,\,\,\,\rm{(wind)}  \\  \,\sin^2(\theta)\,r^{-4}\,t^{-3} \,\,,\,\,\,\,0.1c\leq r/t \leq 0.4c\,\,,\,\,\,\rm{(dyn)}  \\ \,\sin^2(\theta)\,r^{-8}\,t^{-3} \,\,,\,\,\,\,0.4c\leq r/t \leq 0.6c\,\,,\,\,\,\rm{(dyn)}  \end{cases}~
\end{equation}
with the same radial dependence assumed in \citet{Kawaguchi2020} based on numerical-relativity simulations of \citet{Kiuchi2017} and \citet{Hotokezaka2018}, and an angular dependence proposed by \citet{Perego2017} based on fits to numerical-relativity simulations from \citet{Radice2018}. \revised{As in previous works using idealized multi-component ejecta \citep[e.g.][]{Kawaguchi2020}, we assume that the two components are not mixed.} A 3D uniform ($100\times100\times100$) Cartesian grid is used, with a grid resolution of $\Delta\varv=0.0126$c after adding some background material between 0.6 and 0.63c. A density and electron fraction map of the adopted model is shown in Fig.~\ref{fig:ejecta}.

\subsection{Radiative transfer calculations}
\label{sec:sim}

\begin{figure}
\centering
\includegraphics[width=1\columnwidth]{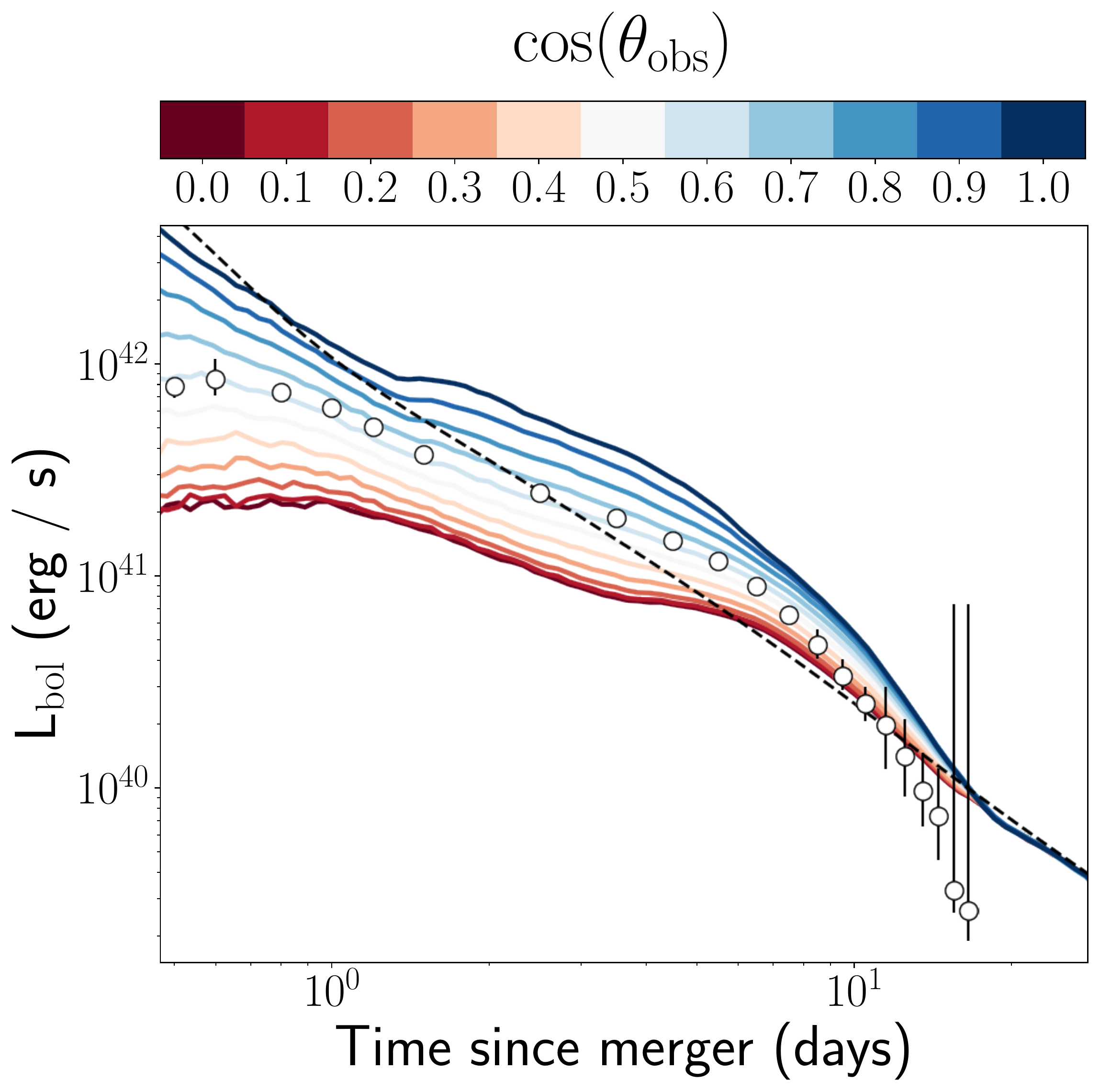}
\caption{Bolometric light curves for the \texttt{FIDUCIAL} model for 11 viewing angles from the merger plane ($\cos\theta_{\rm obs}=0$, edge-on, dark red) to the polar/jet axis ($\cos\theta_{\rm obs}=1$, face-on, dark blue). The black dashed line is the deposition curve for the adopted model, while open circles are bolometric light curves of AT\,2017gfo as inferred in \citet{Waxman2018}.}
\label{fig:lbol_fiducial}
\end{figure}

\begin{figure*}
\centering
\includegraphics[width=1\textwidth]{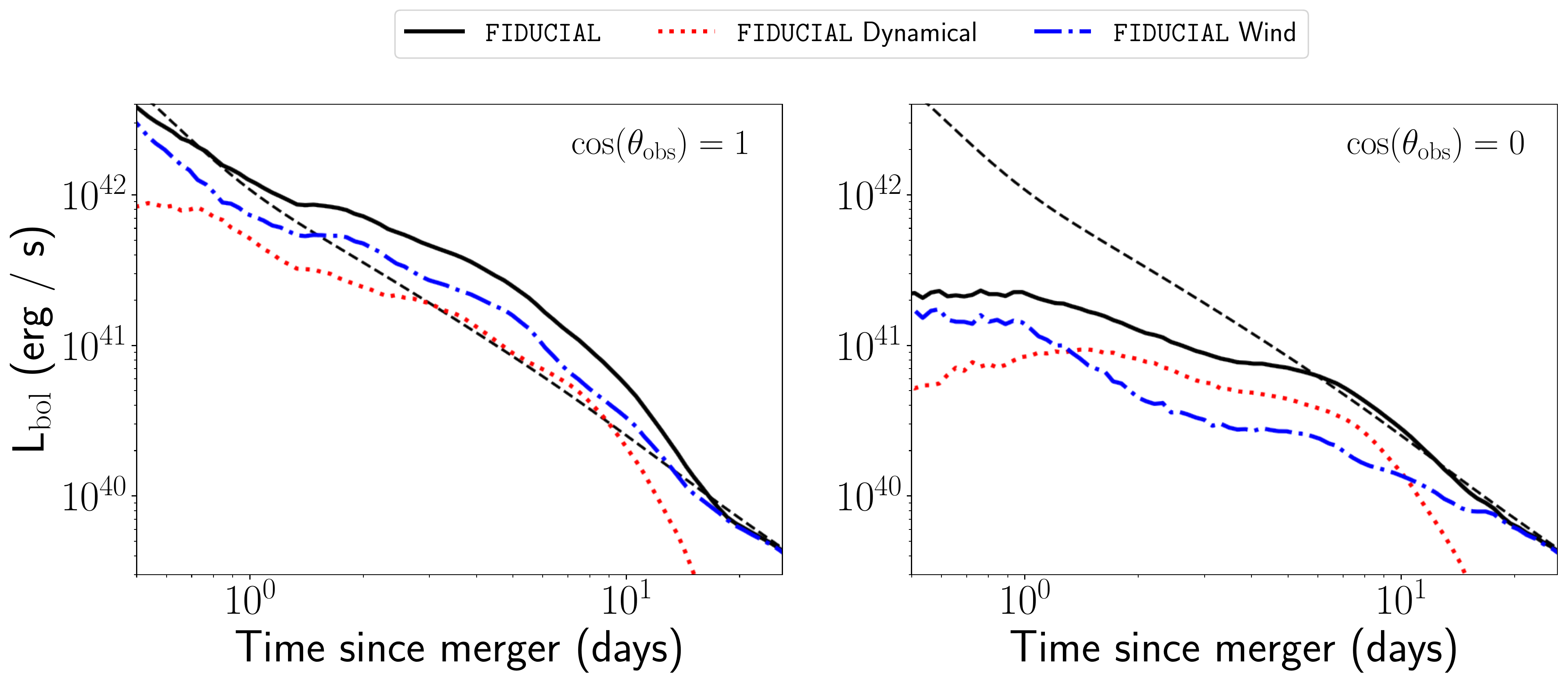}
\caption{Contributions from each ejecta component to the bolometric light curves for the \texttt{FIDUCIAL} model viewed from the polar/jet axis (left panel, $\cos\theta_{\rm obs}=1$, face-on) and merger plane (right panel, $\cos\theta_{\rm obs}=1$, edge-on). The contribution from the dynamical ejecta is shown with dotted red lines, while that from the post-merger disk-wind with dot-dashed blue lines. The resulting final light curves are shown in black and are the same as in Fig.~\ref{fig:lbol_fiducial}.}
\label{fig:lbol_contr}
\end{figure*}

We start from a baseline model using appropriate nuclear heating rates, thermalization efficiencies and opacities as described in Section~\ref{sec:possis}. This model is referred to as \texttt{FIDUCIAL}. We then perform four additional simulations where simplistic assumptions are employed for each or all of these properties. Specifically, the \texttt{SIMPLE-HEAT} model assumes uniform heating rates throughout the ejecta and adopts those from \citet{Korobkin2012} that were tailored to dynamical ejecta with lanthanide-rich compositions (see Equation~\ref{eq:heat}). The \texttt{SIMPLE-THERM} model assumes value of $\epsilon_{\rm th}=0.5$ that is constant with time and uniform throughout the ejecta. The \texttt{SIMPLE-OPAC} model employs grey opacities that are constant with time and take three different values depending on the compositions of the ejecta. Specifically, we choose typical values used in the literature \cite[e.g.][]{Perego2017,Villar2017,Metzger2019,Nicholl2021,Just2022,Collins2022} and set $\kappa_{\rm dyn,lf}=0.5$\,cm$^2$\,g$^{-1}$ and $\kappa_{\rm dyn,lr}=10$\,cm$^2$\,g$^{-1}$ for the lanthanide-free ($Y_{\rm e}\geq0.25$) and lanthanide-rich ($Y_{\rm e}<0.25$) dynamical ejecta, respectively, and $\kappa_{\rm  wind}=3$\,cm$^2$\,g$^{-1}$ for the wind component (the `blue', `red' and `purple' components discussed e.g. in \citealt{Villar2017}). Finally, the \texttt{SIMPLE-ALL} model combines the three assumptions of the \texttt{SIMPLE-HEAT}, \texttt{SIMPLE-THERM} and \texttt{SIMPLE-OPAC} models. The assumptions in each of the five models are summarized in Table~\ref{tab:models}.

For each of the five models described above, we carried out radiative transfer simulations using the new version of \possis~outlined in Section~\ref{sec:possis}. Each simulation adopts a number $N_{\rm ph}=10^7$ of Monte Carlo photon packets, $N_{\rm times}=100$ time-steps from $0.2$ to $30$~d after the merger (logarithmic binning of $\Delta \log t=0.05$), $N_{\rm \lambda}=1000$ wavelength bins from $500$~\AA{} to $10\,\mu$m (logarithmic binning of $\Delta \log\lambda=0.0053$) and $N_{\rm obs}=11$ observer viewing angles equally spaced in cosine ($\Delta\cos\theta_{\rm obs}=0.1$) from the merger plane ($\cos\theta_{\rm obs}=0$, edge-on) to the polar/jet axis ($\cos\theta_{\rm obs}=1$, face-on).

\section{Results}
\label{sec:results}

In this Section, we discuss results of the simulations described in Section~\ref{sec:sim}. We start with the \texttt{FIDUCIAL} model in Section~\ref{sec:fiducial} and then focus on the comparison between the \texttt{FIDUCIAL} and the \texttt{SIMPLE-HEAT}, \texttt{SIMPLE-THERM}, \texttt{SIMPLE-OPAC} and \texttt{SIMPLE-ALL} models in Section~\ref{sec:cf}.

\subsection{Fiducial model}
\label{sec:fiducial}

Focussing on the \texttt{FIDUCIAL} model, we present in the following bolometric light curves (Section~\ref{sec:fidlbol}), broad-band light curves (Section~\ref{sec:fidband}) and spectra (Section~\ref{sec:fidspec}). Although the aim of this work is to study the impact of different assumptions on heating rates, thermalization efficiencies and opacities, rather than fitting data, we show for completeness observations of AT\,2017gfo throughout this section. Specifically, we take bolometric luminosities from \citet{Waxman2018}, broad-band light-curves from \citet{Andreoni2017,Arcavi2017,Chornock2017,
Cowperthwaite2017,Drout2017,Evans2017,
Kasliwal2017,Pian2017,Smartt2017,Tanvir2017,
Troja2017,Utsumi2017,Valenti2017} and X-shooter \citep{Vernet2011} spectra from \citet{Pian2017,Smartt2017}.

\subsubsection{Bolometric light curves}
\label{sec:fidlbol}

\begin{figure}
\centering
\includegraphics[width=1\columnwidth]{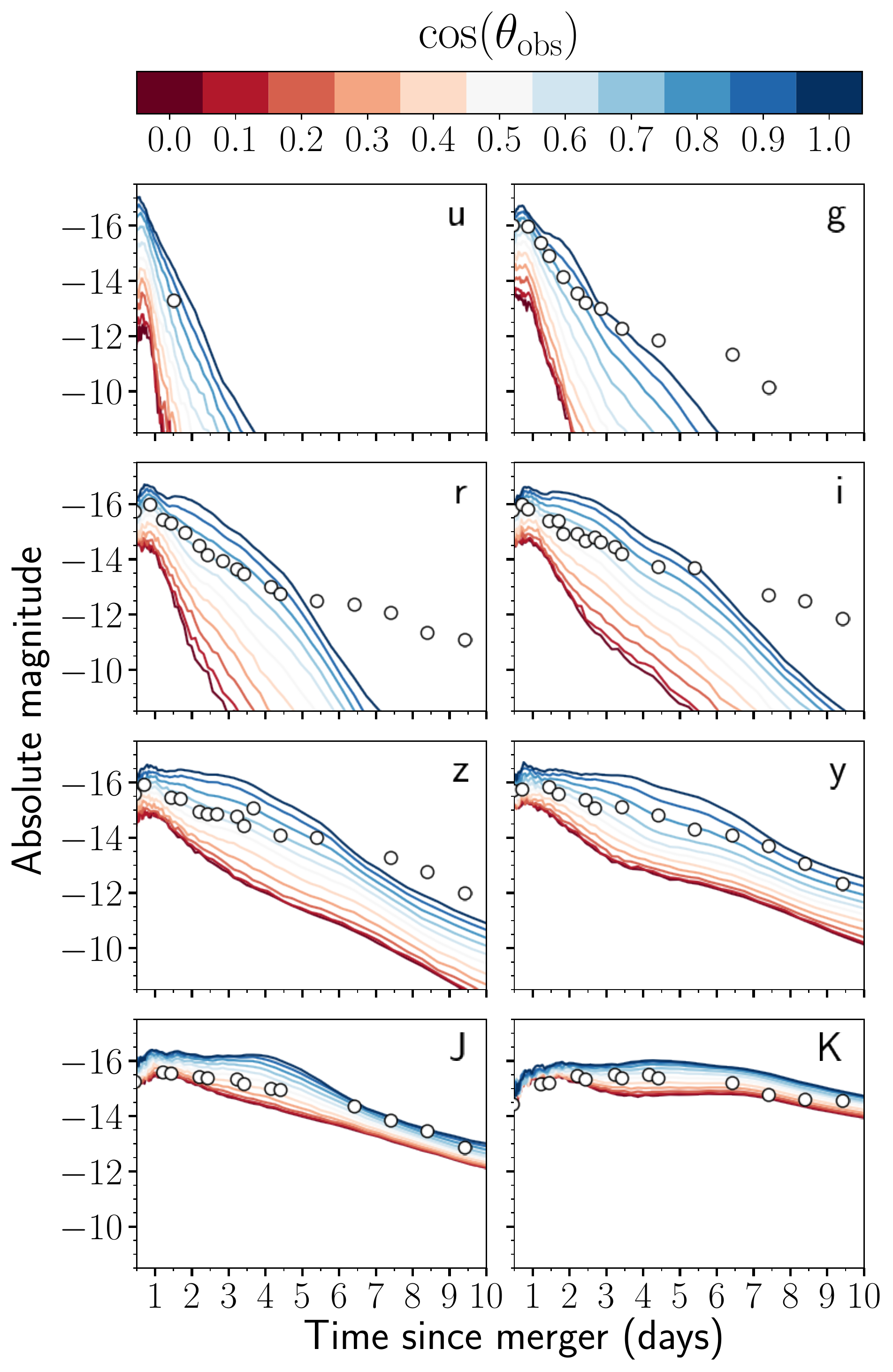}
\caption{Broad-band ugrizyJK light curves for the \texttt{FIDUCIAL} model for 11 viewing angles from the merger plane ($\cos\theta_{\rm obs}=0$, edge-on, dark red) to the polar/jet axis ($\cos\theta_{\rm obs}=1$, face-on, dark blue). Light curves are expressed in absolute magnitude and shown from 0.5 to 10\,d after the merger. Open circles mark observations for AT\,2017gfo, the kilonova associated to GW170817 \citep{Andreoni2017,Arcavi2017,Chornock2017,
Cowperthwaite2017,Drout2017,Evans2017,
Kasliwal2017,Pian2017,Smartt2017,Tanvir2017,
Troja2017,Utsumi2017,Valenti2017}.}
\label{fig:lband_fiducial}
\end{figure}

\begin{figure}
\centering
\includegraphics[width=0.9\columnwidth]{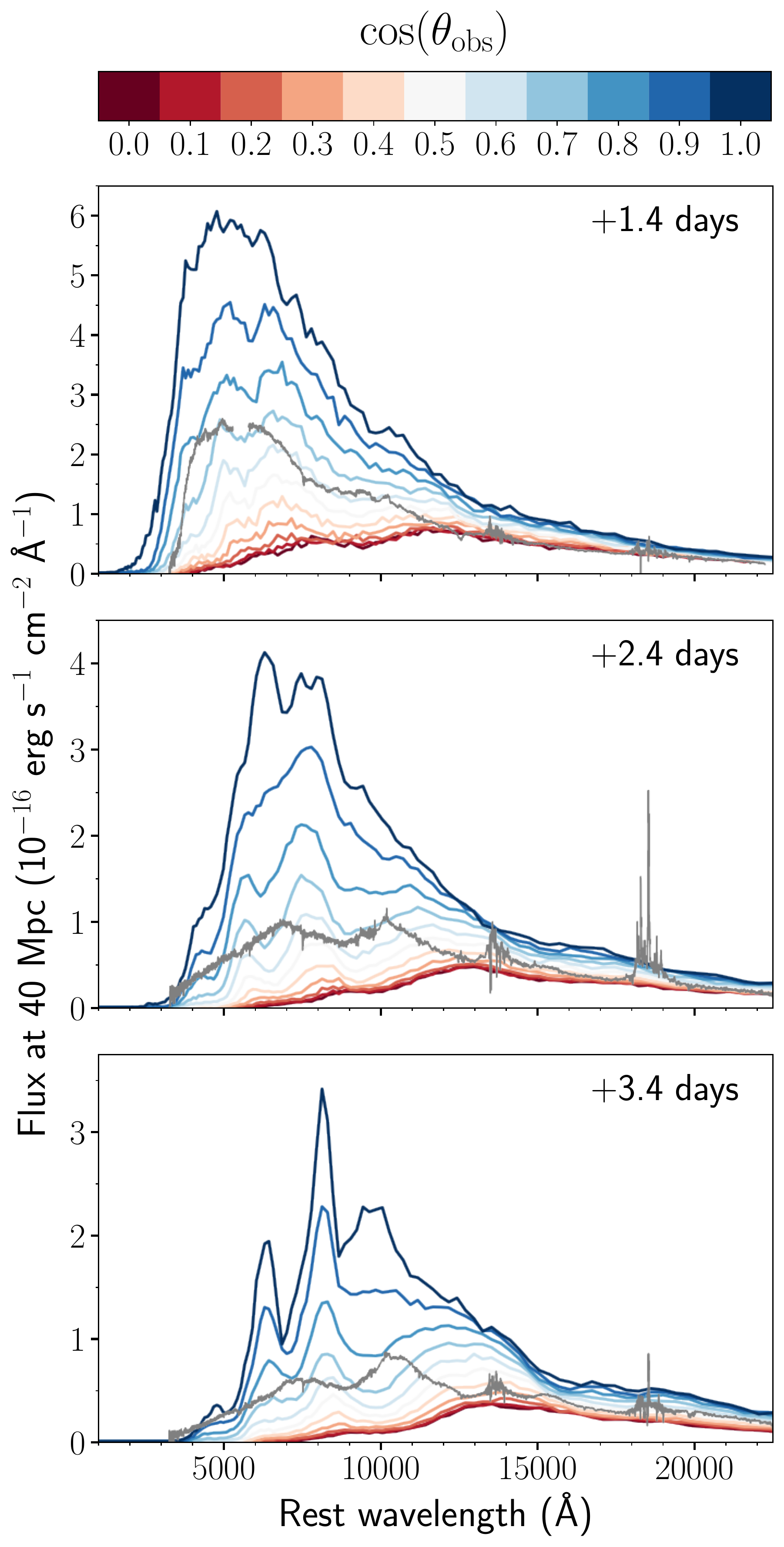}
\caption{Spectra at 1.4, 2.4 and 3.4\,d after the merger (from top to bottom) for the \texttt{FIDUCIAL} model for 11 viewing angles from the merger plane ($\cos\theta_{\rm obs}=0$, edge-on, dark red) to the polar/jet axis ($\cos\theta_{\rm obs}=1$, face-on, dark blue). The epochs are chosen to match those of the dereddened X-shooter spectra of AT\,2017gfo shown in grey \citep[][\revised{regions at $\sim1.4$ and $\sim1.8-1.9\,\mu$m are affected by noise}]{Pian2017,Smartt2017}. Modelled fluxes are re-scaled at 40 Mpc, the inferred distance of GW170817 \citep{Abbott2017a}.}
\label{fig:spec_fiducial}
\end{figure}

\begin{figure*}
\centering
\includegraphics[width=1\textwidth]{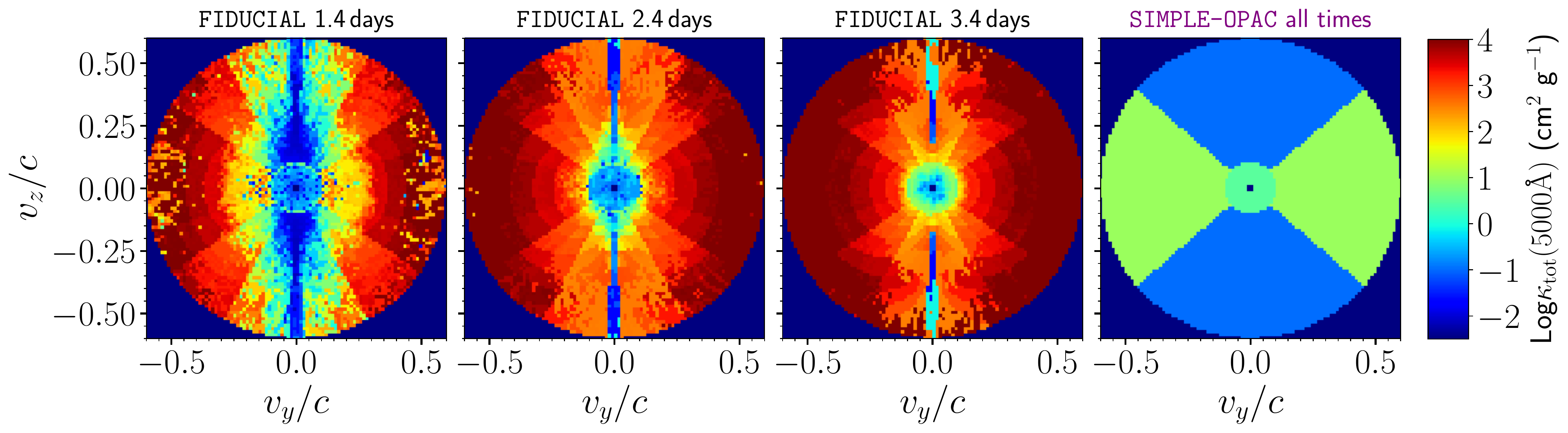}
\caption{Distribution of the opacity $\kappa_{\rm tot}=\kappa_{\rm bb}+\kappa_{\rm sc}$ in the \texttt{FIDUCIAL} and \texttt{SIMPLE-OPAC} models calculated at 5000\,\AA.  The former model adopts opacities from \citet{Tanaka2020} that depend on local values of $\rho$, $T$ and $Y_{\rm e}$, while the latter model assumes grey opacities with $\kappa_{\rm wind}=3$ cm$^2$ g$^{-1}$  for the wind component and $\kappa_{\rm dyn,lf}=0.5$ or $\kappa_{\rm dyn,lr}=10$ cm$^2$ g$^{-1}$ for the lanthanide-poor ($Y_{\rm e}\geq0.25$) and lanthanide-rich ($Y_{\rm e}<0.25$) dynamical ejecta components, respectively. See Section~\ref{sec:opac} for more details. Maps for the \texttt{FIDUCIAL} models are shown at 1.4, 2.4 and 3.4\,d after the merger (same epochs as in Fig.~\ref{fig:spec_fiducial}) while the map for the \texttt{SIMPLE-OPAC} model is the same at all times due to the assumption of grey opacities. The pixelized aspect of the opacities in the \texttt{FIDUCIAL} model is due to the numerical noise in the estimated temperature $T$, see Section~\ref{sec:temp}.}
\label{fig:opac}
\end{figure*}

Bolometric light curves for the \texttt{FIDUCIAL} model are shown in Fig.~\ref{fig:lbol_fiducial} for all viewing angles, while the contributions from the two ejecta components are shown in Fig.~\ref{fig:lbol_contr} for a viewing angle along the jet axis and one in the merger plane. Bolometric light curves do not show a rising phase and continuously decline as found in other works \citep[e.g.][]{Kawaguchi2018,Banerjee2020,Tanaka2020,Collins2022}, indicating that photons can escape early-on from the outermost regions of the ejecta\footnote{\revised{This effect may be due to unreliable and underestimated opacities at the high temperatures ($T\gtrsim20\,000$~K) characterizing ejecta at early phases (M. Tanaka, private communication). However, we note that continuously declining light curves are predicted also for the grey-opacity models in \cite{Collins2022}.}}. As highlighted in Fig.~\ref{fig:lbol_contr}, the early phases ($\lesssim1$\,d) are dominated by flux escaping from the post-merger disk-wind component for all viewing angles. At intermediate phases ($1\lesssim t\lesssim10$\,d), instead, which ejecta component dominates the overall flux depends on the viewing angle: the post-merger disk-wind dominates for an observer along the jet axis (a factor of $\sim2$ higher luminosities than in the dynamical ejecta) while the dynamical ejecta act as a `lanthanide-curtain' \citep{Kasen2015,Wollaeger2018}, obscure the wind ejecta and dominate the total flux for an observer in the merger plane (a factor of $\sim2$ difference in luminosities). Finally, the late phases ($\gtrsim10$\,d) are controlled by the wind ejecta for all orientations. This is a consequence of the wind ejecta being more massive and controlling the deposition curve \revised{(i.e. the total thermalized energy available as a function of time)} at these late phases (see the sudden drop of the dynamical ejecta contribution starting at around $10$\,d).

Despite lacking a peak, light curves are characterized by ondulations and particularly by a sudden change in decay rate \citep[`shoulder',][]{Perez2022,Collins2022} occurring at phases between $\sim1$ and $\sim$10\,d depending on the viewing angle. This marks the transition from the mildly optically-thick regime, when light curves overshoot the deposition curve due to contributions of photons emitted at early epochs, to the optically thin regime, when light curves approach the deposition curve. This transition, and hence the shoulder, occurs at a later phase for viewing angles closer to the merger plane due to the higher opacities and therefore longer diffusion timescales in the dynamical ejecta. We find an additional inflection at earlier times, which is more evident for equatorial viewing angles. 
The viewing angle dependence is relatively strong and decreases with time as the ejecta become optically thin, with a kilonova viewed face-on $\sim6, 3$ and 2 times brighter than one viewed edge-on at $1, 5$ and $10$\,d after the merger, respectively. 

Bolometric luminosities inferred for AT\,2017gfo \citep{Waxman2018} are also reported in Fig.~\ref{fig:lbol_fiducial}. At face value, the model adopted provides a reasonable match to AT\,2017gfo for an observer viewing angle $\sim\theta_{\rm obs}=50^\circ$ ($0.6<\cos\theta_{\rm obs}<0.7$). However, we stress again that here we extract kilonova observables for a single ejecta model and do not attempt to infer ejecta parameters and viewing angle as done elsewhere via model fitting. Nevertheless, the reasonable match found with observations suggests that ejecta properties extracted from numerical-relativity simulations (Section~\ref{sec:ejecta}) can provide a satisfactory description of AT\,2017gfo. This alleviates the claimed tension between numerical-relativity simulations and ejecta parameters inferred for AT\,2017gfo in several studies \citep[e.g.][]{Villar2017} and highlights the importance of using multi-dimensional radiative transfer simulations in place of semi-analytical/one-dimensional kilonova models \citep[see also][]{Kawaguchi2018,Kawaguchi2020,Kedia2022}.

\subsubsection{Broad-band light curves}
\label{sec:fidband}

Fig.~\ref{fig:lband_fiducial} shows $ugrizyJK$ light curves for the \texttt{FIDUCIAL} model. As found in previous works and typical of kilonova light curves, bluer filters are found to peak earlier and decay more rapidly than redder filters. For instance, the $u$-band light curve for a face-on viewing angle peaks at $\sim-17$\,mag in the first day and decline steadily thereafter, with a rate of $\sim3$\,mag d$^{-1}$. In contrast, the infrared $K-$band light curve is longer-lasting and stays at a magnitude of $\sim-16$ for about a week. The light curves show a clear viewing-angle dependence, with a difference between a face-on and edge-on view at 1\,d after the merger of the order of $\sim 4$\,mag in the ultraviolet ($u$), $\sim2$\,mag in optical ($gri$) band and of $\sim 0.5-1.5$\,mag at infrared wavelengths ($zyJK$). 

Photometry of AT\,2017gfo is also reported in Fig.~\ref{fig:lband_fiducial}. The modelled light curves are in reasonable agreement with observations both in terms of brightness and temporal evolution. Similarly to what was seen in bolometric light curves (Section~\ref{sec:fidlbol}), a viewing angle of $\theta_{\rm obs}\sim50^\circ$ provides the best match to the data. While the modelled time-evolution agrees rather well with observations in the infrared bands, models decay too fast in the optical starting from $\sim5-6$\,d after the merger. This discrepancy may be reflective of the adopted ejecta model but also of the limitations of the code at these relatively late epochs. First, the LTE assumption encoded in the adopted opacities \citep{Tanaka2020} is known to break down at these phases \citep{Pognan2022} and might affect the overall light curves in the $\sim5-10$\,d time window that is still moderately optically thick (see Fig.~\ref{fig:lbol_fiducial}). Moreover, the assumption of a perfect coupling between matter and radiation (Section~\ref{sec:temp}) is likely to underestimate the true temperature at these late phases and bias the light curve to redder colors, potentially explaining the observed discrepancy with AT\,2017gfo. Nevertheless, we find a reasonable agreement between model and data that is similar to that found in bolometric light curves and that alleviates the tension in ejecta properties between numerical-relativity simulations and AT\,2017gfo (see discussion in Section~\ref{sec:fidlbol}).

\begin{figure*}
\centering
\includegraphics[width=1\textwidth]{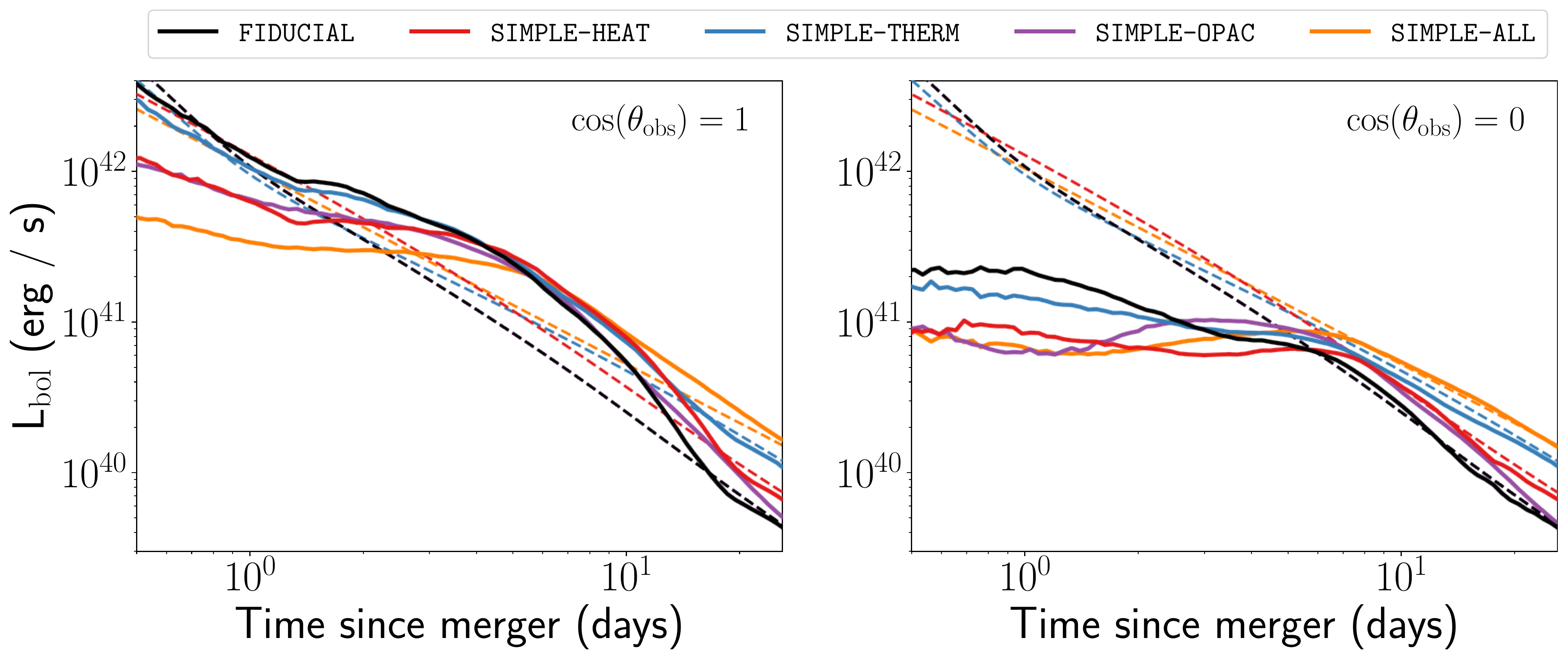}
\caption{Bolometric light curves (solid lines) for the \texttt{FIDUCIAL} model (black) compared to those predicted by the \texttt{SIMPLE-HEAT} (red), \texttt{SIMPLE-THERM} (cyan), \texttt{SIMPLE-OPAC} (purple) and \texttt{SIMPLE-ALL} (orange) models. The left panel shows light curves for an observer along the jet axis (face-on view), while the right panel those for an observer in the merger plane (edge-on view). Dashed lines mark the deposition curves for each model. }
\label{fig:lbol_cf}
\end{figure*}

\begin{figure*}
\centering
\includegraphics[width=0.63\textwidth]{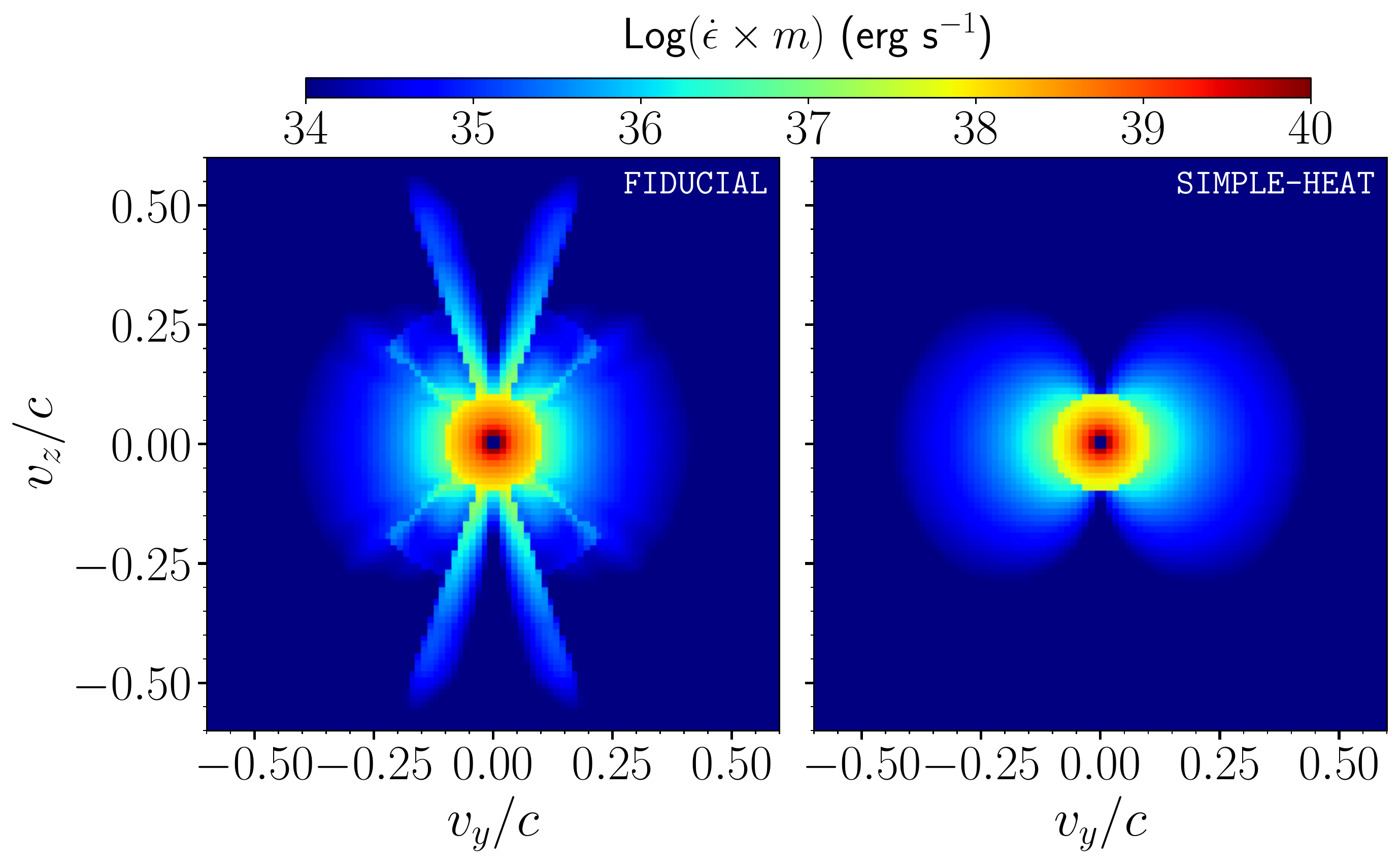}
\includegraphics[width=0.365\textwidth]{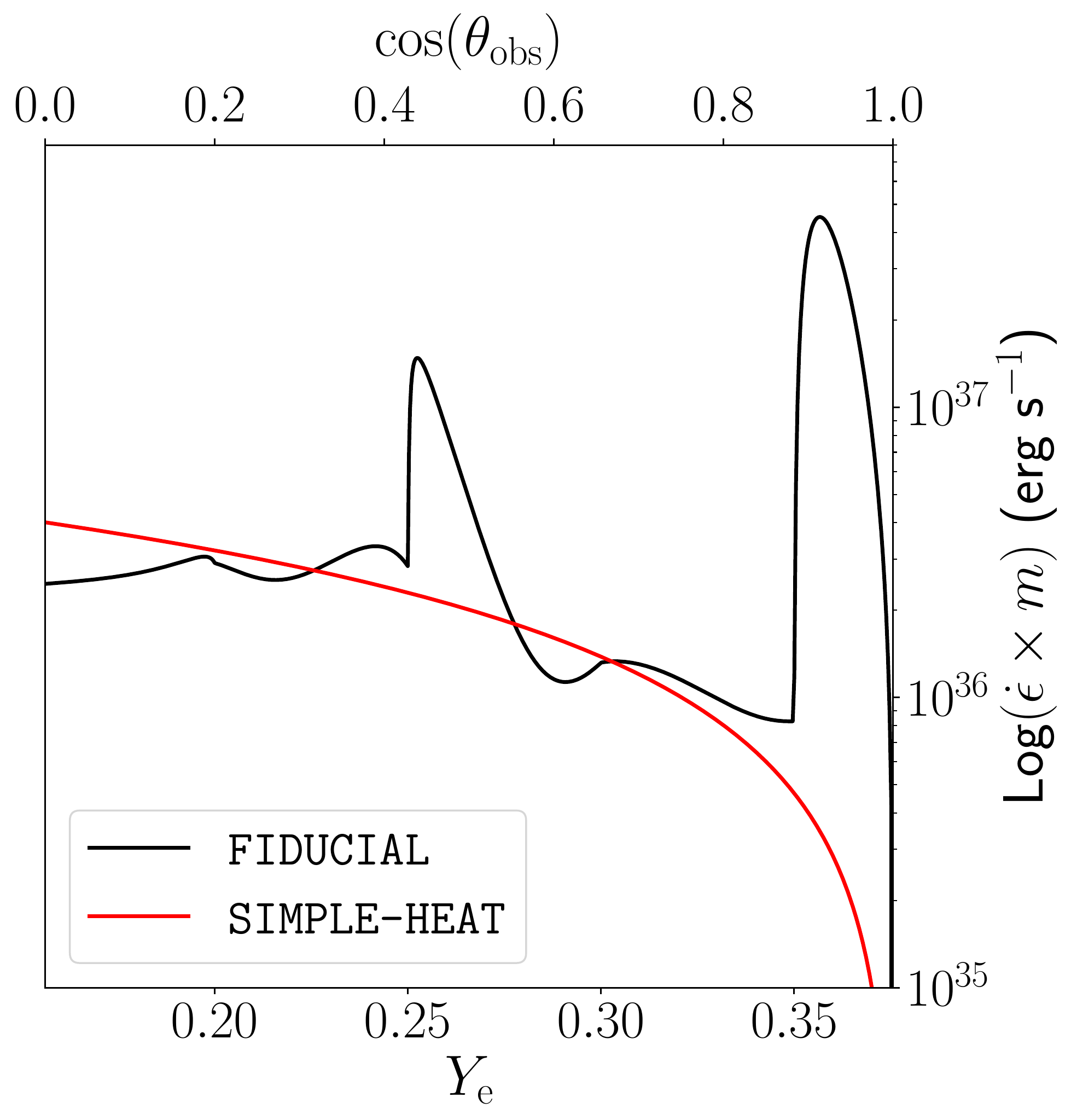}
\caption{Left panels: distribution of $\dot{\epsilon}\times m$ for the \texttt{FIDUCIAL} and the \texttt{SIMPLE-HEAT} model at $0.5$\,d after the merger and for a $\varv_y-\varv_z$ slice in the ejecta. The two models differ in the assumption of the heating rates $\dot{\epsilon}$, with the \texttt{FIDUCIAL} model using $Y_{\rm e}-$ and $\varv-$dependent rates from \citet{Rosswog2022} and the \texttt{SIMPLE-HEAT} model values from \citet{Korobkin2012} throughout the entire ejecta. Right panel: $\dot{\epsilon}\times m$ at $0.5$\,d and at a velocity $\varv=0.25$c (dynamical ejecta) as a function of the viewing angle $\theta_{\rm obs}$ or equivalently $Y_{\rm e}$ (see eq.~\ref{eq:ye}). The curves are computed using the analytic prescriptions described in Section~\ref{sec:heat} and Section~\ref{sec:ejecta} rather than from the maps in the left and middle panels. The thermalization efficiencies $\epsilon_{\rm therm}$ are not considered here as they are the same in both the \texttt{FIDUCIAL} and the \texttt{SIMPLE-HEAT} models.}
\label{fig:doteps}
\end{figure*}

\subsubsection{Spectra}
\label{sec:fidspec}

Spectra at 1.4, 2.4 and 3.4\,d after the merger are shown in Fig.~\ref{fig:spec_fiducial} for the 11 viewing angles. The epochs are chosen to match those of the first three spectra of AT\,2017gfo taken with X-shooter \citep{Pian2017,Smartt2017}, which are also reported in the same figure for comparison. Spectra for viewing angles close to the merger plane are fainter and redder than those for viewing angles close to the jet axis, with most of the flux escaping at infrared wavelengths. Moreover, the time evolution reflects the color evolution discussed in Section~\ref{sec:fidband}, with a rapid reddening of the spectra observed for all the viewing angles. These effects are a consequence of the increase in opacities moving towards edge-on orientations and of their rapid increase with time, see the three panels on the left in Fig.~\ref{fig:opac}. 

While spectra in the first version of \possis~\citep{Bulla2019b} were smooth continua due to the simplified assumption on the opacities, the new version with more sophisticated opacities \citep{Tanaka2020} produces spectra with broad features that develop rapidly in strength from 1.4 to 3.4\,d after the merger. 

Once again (see Sections~\ref{sec:fidlbol}-\ref{sec:fidband}), a good agreement is found between observed and modelled spectra for polar-to-intermediate viewing angles. The model suggests the presence of a broad feature around $1\mu$m that could potentially be associated with the one in AT\,2017gfo $\sim8000$\,\AA{} and ascribed to Sr II \citep{Watson2019,Domoto2021}. However, the opacity table used from \citet{Tanaka2020} does not include information about atomic processes responsible for individual transitions and therefore we can not perform such an association. Moreover, the opacities from \citet{Tanaka2020} are not calibrated experimentally and therefore have relatively large uncertainties on the wavelength of each transitions, thus preventing us to make a one-to-one comparison between modelled and observed features.



\subsection{Simplistic models}
\label{sec:cf}

In this Section we present the comparisons between the \texttt{FIDUCIAL} model and the \texttt{SIMPLE-HEAT}, \texttt{SIMPLE-THERM}, \texttt{SIMPLE-OPAC} and \texttt{SIMPLE-ALL} models in terms of bolometric light curves (Section~\ref{sec:simplbol}), broad-band light curves (Section~\ref{sec:simplband}) and spectra (Section~\ref{sec:simplspec}). 

\subsubsection{Bolometric light curves}
\label{sec:simplbol}

\begin{figure*}
\centering
\includegraphics[width=0.98\textwidth]{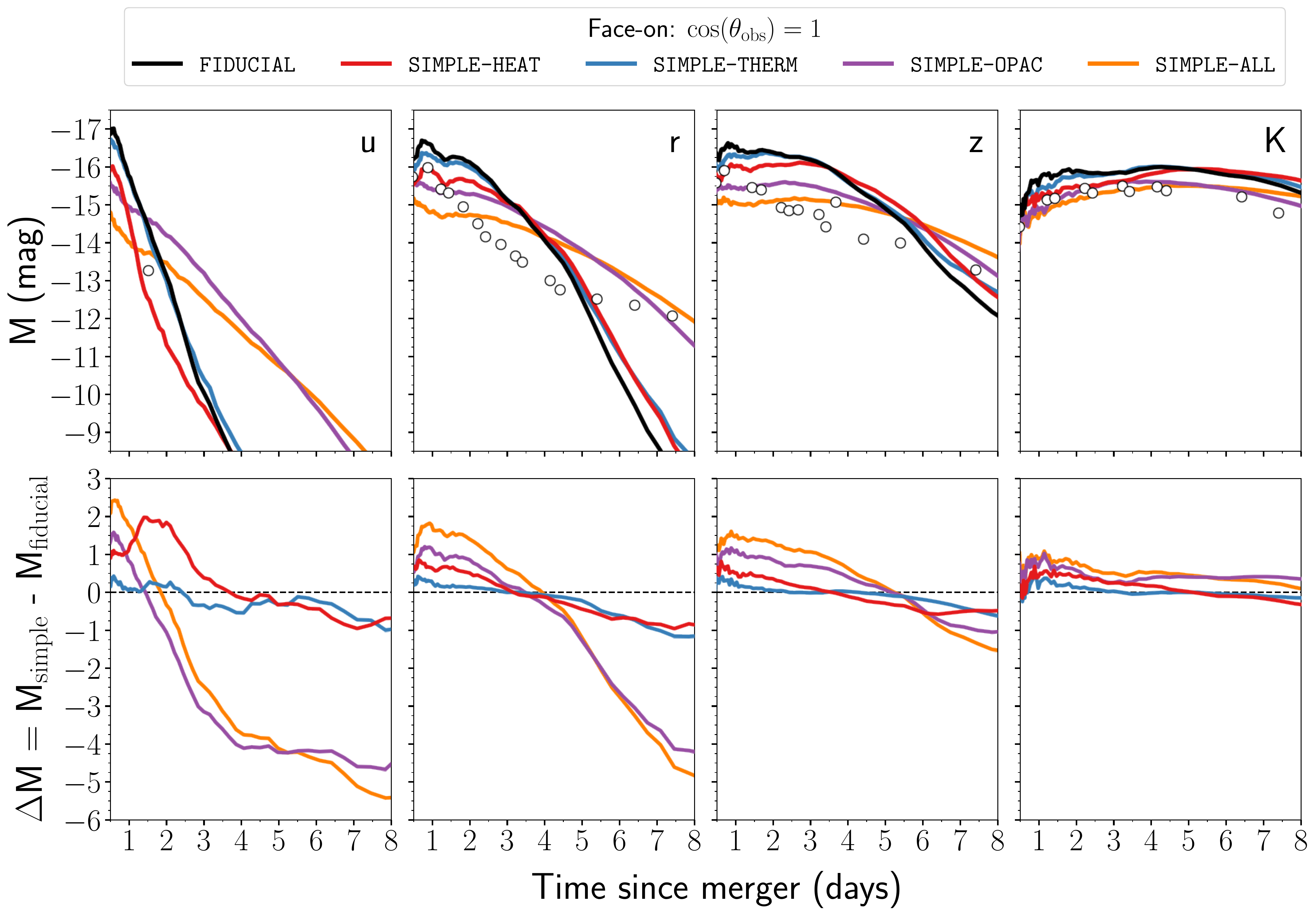}
\caption{Broad-band light curves in the $u$, $r$, $z$ and $K$ filters for a face-on observer ($\cos\theta_{\rm obs}=1$) and for all the five models (top panels). Deviations from the \texttt{FIDUCIAL} model are shown in the lower panels for the four \texttt{SIMPLE-} models. \revised{As in Fig.~\ref{fig:lband_fiducial}, open circles mark observations for AT\,2017gfo, the kilonova associated to GW170817.}}
\label{fig:lband_faceon_cf}
\end{figure*}

\begin{figure*}
\centering
\includegraphics[width=0.98\textwidth]{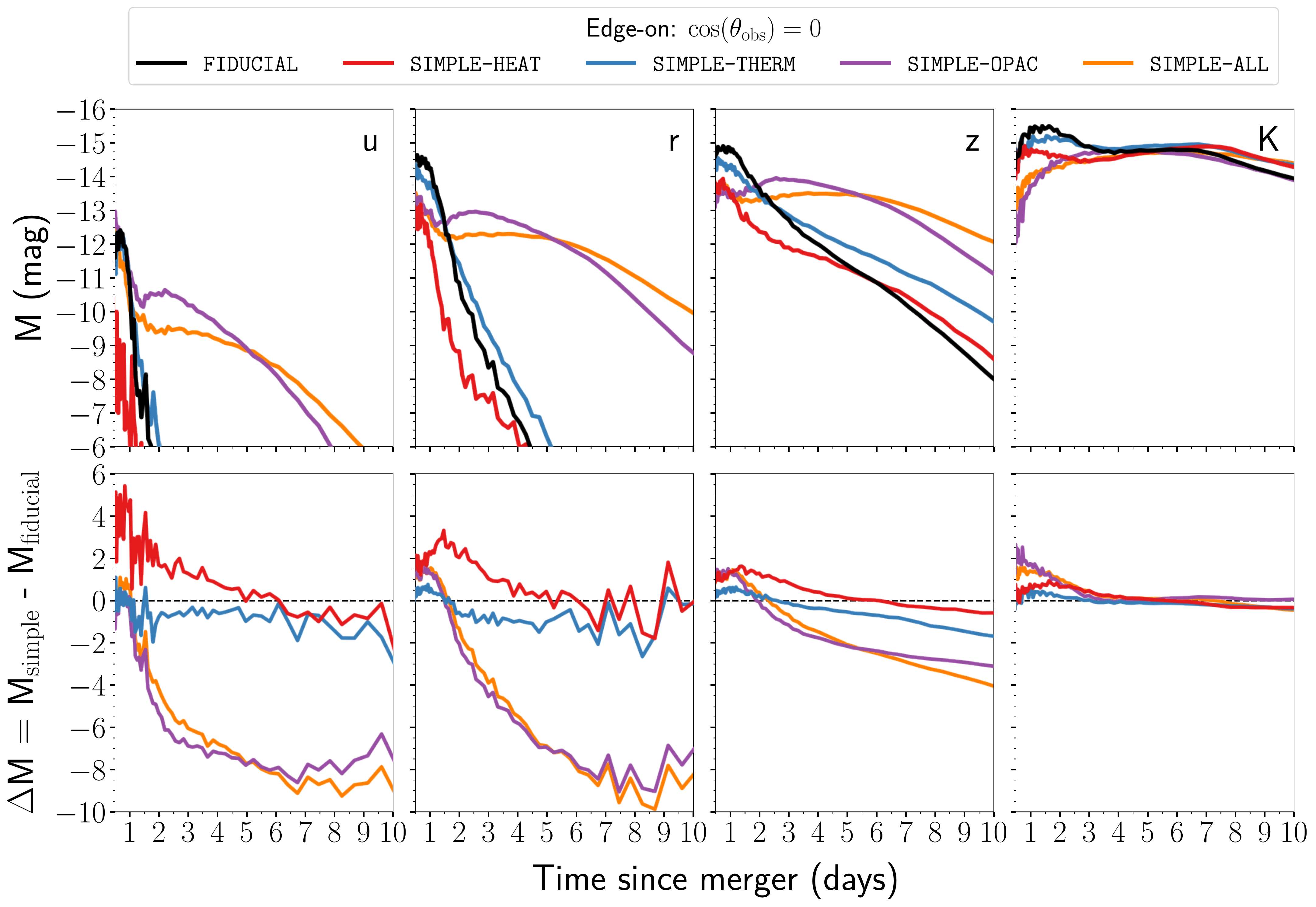}
\caption{Same as Fig.~\ref{fig:lband_faceon_cf} but for an edge-on observer ($\cos\theta_{\rm obs}=0$).}
\label{fig:lband_edgeon_cf}
\end{figure*}

The comparison between bolometric light curves for all the five models is shown in Fig.~\ref{fig:lbol_cf} for an observer along the jet axis and one in the merger plane. While light curves are relatively similar about a week after the merger, clear departures from the \texttt{FIDUCIAL} model are seen for all the \texttt{SIMPLE-} models.

The \texttt{SIMPLE-HEAT} model is fainter than the \texttt{FIDUCIAL} model at $\lesssim5$\,d since merger, while brighter thereafter. Focusing first on the early phases, the discrepancy can be understood by looking at the quantity $\dot{\epsilon} \times m$, where $m$ is the mass, for the two models as shown in Fig.~\ref{fig:doteps} for a time of 0.5\,d. In the \texttt{SIMPLE-HEAT} model, heating rates are taken from \citep{Korobkin2012} and assumed to be uniform throughout the ejecta. Therefore, $\dot{\epsilon} \times m$ tracks the density/mass distribution within the ejecta, as shown e.g. in the right panel of Fig.~\ref{fig:doteps} for a shell at $\varv=0.25$c in the dynamical ejecta (see red line). In contrast, the \texttt{FIDUCIAL} model assumes $Y_{\rm e}-$ and $\varv-$dependent heating rates from \citet{Rosswog2022} and thus leads to a more complex structure of $\dot{\epsilon} \times m$ within the ejecta (see left panel). In particular, a fair fraction of the energy is injected in regions of the dynamical ejecta with $0.25\lesssim Y_{\rm e}\lesssim0.27$ and $0.35\lesssim Y_{\rm e}\lesssim0.38$ (see stripes in the left panel and ``bumps'' in the right panel). As a result, radiation for the \texttt{FIDUCIAL} model is created in more optically thin regions than it is the case for the \texttt{SIMPLE-HEAT} model, hence leading to brighter light curves at early phases. At the same time, this leads to shorter diffusion time-scales thus explaining the behaviour at late phases in comparison to the \texttt{SIMPLE-HEAT} model. 

The \texttt{SIMPLE-THERM} model is the one that is closer to the \texttt{FIDUCIAL} model of all the cases considered. The predicted light-curve difference appears to reflect the difference in deposition curves between the two models. Specifically, assuming $\epsilon_{\rm therm}=0.5$ throughout the whole ejecta underestimates the true amount of thermal energy deposited at relatively early times and overestimates the same at late times. The corresponding transition occurs at $\sim2$\,d after the merger for both orientations, which is the same phase at which the light curves for the \texttt{SIMPLE-THERM} model switch from being fainter to being brighter than those for the \texttt{FIDUCIAL} model.

The \texttt{SIMPLE-OPAC} model shows a qualitatively similar behaviour to those found for the \texttt{SIMPLE-HEAT} and \texttt{SIMPLE-THERM} models, in that it is fainter than the \texttt{FIDUCIAL} model at early phases and brighter at later epochs. This discrepancy is a direct consequence of the grey opacities assumed in the simplistic model. Fig.~\ref{fig:opac} compares the opacity values assumed by the two models at 1.4, 2.4 and 3.4\,d after the merger and at 5000\,\AA, wavelength at which most of the emission comes from at these phases (see Fig.~\ref{fig:spec_fiducial}). The grey opacities assumed overestimate the opacity values from \citet{Tanaka2020} for the disk-wind component, from which most of the flux is coming from at these phases (Fig.~\ref{fig:lbol_contr}). In addition, bound-bound opacities in the wind ejecta rapidly increase with time (cf three panels in Fig.~\ref{fig:opac}) due to temperature decreasing and matter recombining, and eventually become larger than those assumed by the grey-opacity \texttt{SIMPLE-OPAC} model. As a consequence, photons travelling towards polar regions take longer to diffuse out in the \texttt{SIMPLE-OPAC} compared to the \texttt{FIDUCIAL} model, leading to fainter and longer-lasting light curves (see left panel of Fig.~\ref{fig:lbol_cf}). A similar effect is found for an edge-on view of the system (right panel) in the first day after the merger when the emitted light is dominated by the post-merger disk-wind, as shown in Fig.~\ref{fig:lbol_contr}. Between 1 and 10\,d, however, the kilonova emission is dominated by the dynamical-ejecta component (Fig.~\ref{fig:lbol_contr}), where opacities are greatly underestimated in the \texttt{SIMPLE-OPAC} compared to \texttt{FIDUCIAL} model. Therefore, the latter is generally brighter than the former at these phases.

The \texttt{SIMPLE-ALL} model is the one with the largest deviations from the \texttt{FIDUCIAL} model. Following the discussions above, the lower luminosities at early times are naturally driven by the simplistic assumptions on heating rates and opacity, while the higher luminosities at late times by the simplistic assumption on heating rates and thermalization efficiencies (see different in deposition curves at $\gtrsim10$\,d after the merger in Fig.~\ref{fig:lbol_cf}).

\begin{figure*}
\centering
\includegraphics[width=0.85\textwidth]{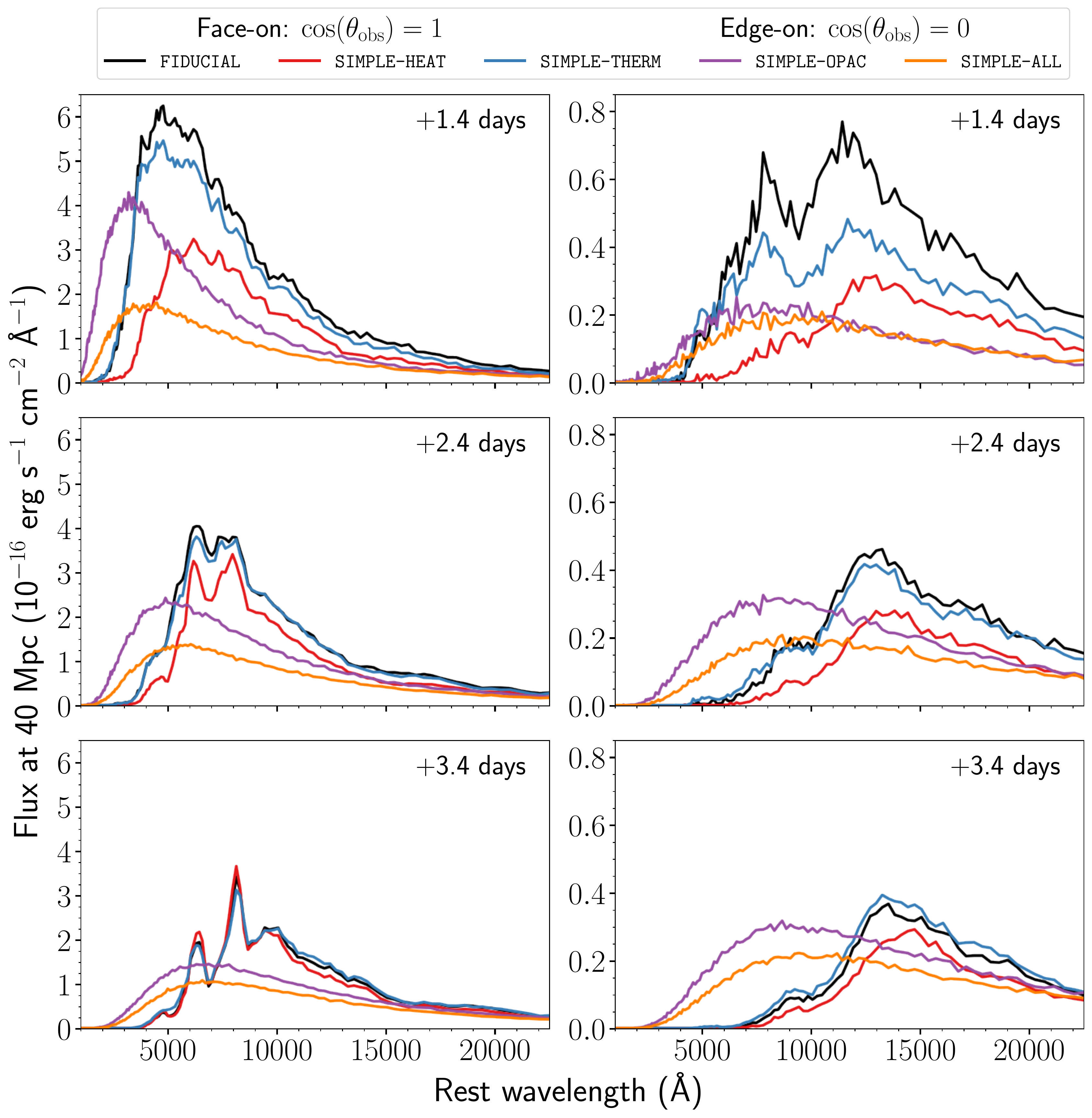}
\caption{Spectra at 1.4, 2.4 and 3.4\,d after the merger (from top to bottom) for all the five models and for a face-on ($\cos\theta_{\rm obs}=1$, left) and edge-on ($\cos\theta_{\rm obs}=0$, right) observer. As in Fig.~\ref{fig:spec_fiducial}, fluxes are scaled to $40$\,Mpc \citep{Abbott2017a} and epochs chosen to match those of the X-shooter spectra of AT\,2017gfo \citep{Pian2017,Smartt2017}. Note that the y-scale is different between the left and right panels. \revised{Spectra for the \texttt{SIMPLE-OPAC} and \texttt{SIMPLE-ALL} models are consistent with smooth continua because of the grey-opacity assumption, with small deviations due to Monte Carlo numerical noise.}}
\label{fig:spec_cf}
\end{figure*}

\subsubsection{Broad-band light curves}
\label{sec:simplband}

Light curves in the $u$, $r$, $z$ and $K$ bands are shown for all the models in Fig.~\ref{fig:lband_faceon_cf} (face-on view) and Fig.~\ref{fig:lband_edgeon_cf} (edge-on view). The trend observed in the bolometric light curves (Section~\ref{sec:simplbol}) are confirmed in the broad-band light curves, with the simplistic models generally fainter than the \texttt{FIDUCIAL} model at early phases and brighter at later phases. Moreover, the \texttt{SIMPLE-THERM} model is the one in closer agreement with the \texttt{FIDUCIAL} model, followed in order by the \texttt{SIMPLE-HEAT}, the \texttt{SIMPLE-OPAC} and the \texttt{SIMPLE-ALL} models. Finally, the discrepancies generally decrease moving to longer wavelengths. This effect can be ascribed to the reduced optical depths moving to the infrared and therefore the relatively smaller sensitivity to the simplistic assumptions on opacities and heating rates (see discussion in Section~\ref{sec:simplbol}).

For a face-on observer (Fig.~\ref{fig:lband_faceon_cf}), the \texttt{SIMPLE-THERM} model agrees relatively well with the \texttt{FIDUCIAL} model in all bands and for the first $\sim$ week after the merger, with deviations that are typically $|\Delta M|\lesssim0.5$\,mag. The \texttt{SIMPLE-HEAT} model reaches $|\Delta M|\sim2$\,mag in the $u-$band and $|\Delta M|\sim0.6$\,mag at longer wavelengths a couple of days after the merger. Discrepancies for these two models generally increase in the bluer bands starting from $\sim6-7$\,d after the merger, but are still restricted to $|\Delta M|\lesssim1$\,mag. The other two models, \texttt{SIMPLE-OPAC} and \texttt{SIMPLE-ALL}, show larger discrepancies from the \texttt{FIDUCIAL} model. In the first couple of days after the merger, deviations reach a maximum around $|\Delta M|\sim1-1.5$\,mag for the former and $|\Delta M|\sim1-2.5$\,mag for the latter model, with larger values in the $u-$band and smaller values in the $K-$band. At later times, ultraviolet ($u$) and optical ($r$) filters can become $|\Delta M|\sim4-5$\,mag brighter than the \texttt{FIDUCIAL} model, while $z$ and $K$ light curves have similar deviations to those at early times.  \revised{We note that a better match to the late-time ($\gtrsim 6$\,d) $r-$band light curve of AT\,2017gfo is found in the \texttt{SIMPLE-OPAC} and \texttt{SIMPLE-ALL} compared to the \texttt{FIDUCIAL} model, which highlights possible misleading agreements for this simplified grey-opacity models.}

The trends predicted for an edge-on observer (Fig.~\ref{fig:lband_edgeon_cf}) are qualitatively similar to those for a face-on observer although with typically larger discrepancies. Specifically, deviations from the \texttt{FIDUCIAL} model are about twice as large as those found for a face-on observer. For instance, discrepancies in ultraviolet and optical filters are in the range $-1.5\lesssim\Delta M\lesssim0.5$\,mag for the \texttt{SIMPLE-THERM} model, $-1\lesssim\Delta M\lesssim4$\,mag for the \texttt{SIMPLE-HEAT} model and $-10\lesssim\Delta M\lesssim1.5$\,mag for the \texttt{SIMPLE-OPAC} and  \texttt{SIMPLE-ALL} models.

\subsubsection{Spectra}
\label{sec:simplspec}

Spectra at 1.4, 2.4 and 3.4\,d after the merger are shown in Fig.~\ref{fig:spec_cf} for all the five models discussed in this work. At these early phases, all the four \texttt{SIMPLE-} models are fainter than the \texttt{FIDUCIAL} models as evidenced in the light curves (see Sections~\ref{sec:simplbol}-\ref{sec:simplband}). The spectral shape of the \texttt{SIMPLE-THERM} is relatively similar to the one in the \texttt{FIDUCIAL} model as the temperatures are only slightly affected by the difference in the thermalization efficiencies at these early phases. Spectra for the \texttt{SIMPLE-HEAT} model peak at slightly longer wavelengths (i.e. they are cooler) due to the overall lower heating in this model (see right panel of Fig.~\ref{fig:doteps}). The \texttt{SIMPLE-OPAC} and the \texttt{SIMPLE-ALL} models, instead, have spectra peaking at shorter wavelengths, which is a natural consequence of the difference in opacities compared to the \texttt{FIDUCIAL} model. As shown in Fig.~\ref{fig:opac}, these models assume grey opacities and tend to underestimate the opacities of the \texttt{FIDUCIAL} model at the three phases considered. Radiation can therefore escape more easily at short wavelengths and suffer less reprocessing to longer wavelengths than it is the case in the \texttt{FIDUCIAL} model (see Section~\ref{sec:fidspec}). We note that spectra in the \texttt{SIMPLE-OPAC} and \texttt{SIMPLE-ALL} models are smooth continua due to the grey-opacity assumptions \revised{(with deviations cause by Monte Carlo numerical noise)}, while spectra for the other three models using wavelength-dependent opacities from \citet{Tanaka2020} are characterized by spectral features.

\section{Conclusions}
\label{sec:conclusions}

We have presented an improved version of the 3-D Monte Carlo radiative code \possis. Compared to the first version of the code \citep{Bulla2019b}, the new \possis~calculates temperature from the mean intensity of the radiation field and use nuclear heating rates, thermalization efficiencies and opacities as a function of local properties of the ejecta. Specifically, heating rates as a function of velocity and electron fraction are taken from \citet{Rosswog2022}, density-dependent thermalization efficiencies are computed accounting for the contribution of each species ($\alpha-$, $\beta-$, $\gamma-$ particles and fission fragments) following \citet{Barnes2016} and \citet{Wollaeger2018}, while opacities as a function of time, wavelength, density, temperature and electron fraction are taken from \citet{Tanaka2020}.

With the new version of the code, we computed kilonova spectra and light curves for an axially-symmetric \texttt{FIDUCIAL} model with parameters guided from numerical-relativity simulations of neutron star mergers \citep{Radice2018,Nedora2021}. The grid includes two components: a dynamical ejecta component with $m_{\rm dyn}=0.005\,M_\odot$, expanding at an  average velocity of $\bar{\varv}_{\rm dyn}=0.2$c and with an angular-dependent electron fraction with average value of $\bar{Y}_{\rm e, dyn}=0.2$, and a spherical post-merger disk-wind component with $m_{\rm wind}=0.05\,M_\odot$, average velocity $\bar{\varv}_{\rm wind}=0.05$c and electron fraction $Y_{\rm e, dyn}=0.3$. Although the aim of this work is not to perform model fitting and parameter inference, we find a reasonably good match between the \texttt{FIDUCIAL} model and AT\,2017gfo. This alleviates the claimed tension between ejecta parameters predicted from numerical-relativity simulations and those inferred for AT\,2017gfo using semi-analytical codes \citep[e.g.][]{Villar2017}, highlighting the importance of multi-dimensional radiative transfer simulations for kilonova modelling. 

We then investigated the critical role of nuclear heating rates, thermalization efficiencies and opacities in shaping the spectra and light curves of kilonovae by performing four additional models: the \texttt{SIMPLE-HEAT} model where heating rates are assumed to be uniform throughout the ejecta \citep{Korobkin2012}, the \texttt{SIMPLE-THERM} model where the thermalization efficiency is set to 0.5 at all times and throughout the entire ejecta, the \texttt{SIMPLE-OPAC} model where grey opacities are assumed for three different ejecta components (``red'', ``blue'' and ``purple'', \citealt{Villar2017}), and the \texttt{SIMPLE-ALL} model where all the three simplistic assumptions are combined. We find large deviations from the \texttt{FIDUCIAL} model for all the four models considered. These increase going from the \texttt{SIMPLE-THERM} to the \texttt{SIMPLE-HEAT}, to the \texttt{SIMPLE-OPAC} and \texttt{SIMPLE-ALL} models and are in the range of $1-10$\,mag depending on the assumed viewing angle (larger deviations for edge-on compared to face-on observers), time (deviations increasing with time) and filter (deviations decreasing with wavelengths). 

Our study suggests that kilonova models adopting one or multiple of the simplistic assumptions discussed in this work ought to be treated with caution and that appropriate error bars ought to be added to kilonova light curves when performing parameter inference. To what extent these simplified assumptions affect parameter inference in large kilonova grids remains to be seen and will be investigated in future works.

\section*{Acknowledgements}

I thank Christine Collins, Tim Dietrich, Oleg Korobkin, Mark Magee, Stephan Rosswog and Stuart Sim for fruitful discussions. I am grateful to Masaomi Tanaka for kindly sharing the set of time- and wavelength-dependent opacities used in this work. This work was supported by the European Union’s Horizon 2020 Programme under the AHEAD2020 project (grant agreement n. 871158) and by the National Science Foundation under Grant No. PHY-1430152 (JINA Center for the Evolution of the Elements). The simulations were performed on resources provided by the Swedish National Infrastructure for Computing (SNIC) at Kebnekaise partially funded by the Swedish Research Council through grant agreement no. 2018-05973. \revised{I thank the anonymous referee for constructive comments.}

\section*{Data Availability}

The simulations performed in this study will be made publicly available at \url{https://github.com/mbulla/kilonova_models}.



\bibliographystyle{mnras}
\bibliography{astrobulla} 







\bsp	
\label{lastpage}
\end{document}